\documentclass[conference]{IEEEtran}
\IEEEoverridecommandlockouts
\usepackage{cite}
\usepackage{amsmath,amssymb,amsfonts}
\usepackage{algorithmic}
\usepackage{graphicx}
\usepackage{textcomp}
\usepackage{xcolor}
\usepackage{comment}
\usepackage{url}
\usepackage{hyperref}
\hypersetup{colorlinks=true,allcolors=black}
\usepackage{multirow}
\usepackage{acronym}
\usepackage{subfigure}

\acrodef{dl}[DL]{Deep Learning}
\acrodef{ml}[ML]{Machine Learning}
\acrodef{cpu}[CPU]{Central Processing Unit}
\acrodef{pmu}[PMU]{Performance Monitoring Unit}
\acrodef{pmu}[PMU]{Performance Monitoring Unit}
\acrodef{pmc}[PMC]{Performance Monitoring Counter}
\acrodef{dram}[DRAM]{Dynamic Random Access Memory}
\acrodef{mpsoc}[MPSoC]{Multi-Processor System-on-Chip}

\acrodef{ros2}[ROS2]{Robot Operating System 2}
\acrodef{rt}[RT]{Real-Time}
\acrodef{nrt}[nRT]{non-Real-Time}
\acrodef{sdvbs}[SD-VBS]{San Diego Vision Benchmark Suite}
\acrodef{soc}[SoC]{System-on-Chip}
\acrodef{qos}[QoS]{Quality of Service}

\acrodef{cots}[COTS]{Commercial Off-The-Shelf}
\acrodef{os}[OS]{Operating System}
\acrodef{hpec}[HPEC]{High Performance Embedded Computer}
\acrodef{wcet}[WCET]{Worst Case Execution Time}
\acrodef{dds}[DDS]{Data Distribution Service}
\acrodef{gpu}[GPU]{Graphics Processing Unit}

\acrodef{ct}[CT]{Critical Task}
\acrodef{nct}[nCT]{Non-Critical Task}

\acrodef{adas}[ADAS]{Advanced Driver Assistance Systems}
\acrodef{ad}[AD]{Autonomous Driving}

\acrodef{cnn}[CNN]{Convolutional Neural Network}
\acrodef{dnn}[DNN]{Deep Neural Network}

\graphicspath{{../pdf/}{../jpeg/}{2_plots/}}
\def\BibTeX{{\rm B\kern-.05em{\sc i\kern-.025em b}\kern-.08em
    T\kern-.1667em\lower.7ex\hbox{E}\kern-.125emX}}

\usepackage[T1]{fontenc}
\definecolor{ForestGreen}{rgb}{0.0, 0.27, 0.13}

\begin{document}

\title{ROSGuard: A Bandwidth Regulation Mechanism\\ for ROS2-based Applications
\thanks{This work has received funding from the European Union’s Horizon Europe Programme under the SAFEXPLAIN Project (www.safexplain.eu), grant agreement num. 101069595, and from the Spanish Ministry of Science and Innovation under project PLEC2023-010240, CAPSULIA.}
}

\author{
\IEEEauthorblockN{Jon Altonaga Puente}
\IEEEauthorblockA{\textit{Cybersecurity and Dependability} \\
\textit{Ikerlan}\\
Arrasate, Spain \\
jaltonaga@ikerlan.es}
\and
\IEEEauthorblockN{Enrico Mezzetti}
\IEEEauthorblockA{\textit{High-Performance Embedded Systems} \\
\textit{Barcelona Supercomputing Center}\\
Barcelona, Spain \\
enrico.mezzetti@bsc.es}
\and
\IEEEauthorblockN{Irune Agirre Troncoso}
\IEEEauthorblockA{\textit{Cybersecurity and Dependability} \\
\textit{Ikerlan}\\
Arrasate, Spain \\
iagirre@ikerlan.es}
\and
\IEEEauthorblockN{Jaume Abella Ferrer}
\IEEEauthorblockA{\textit{High-Performance Embedded Systems} \\
\textit{Barcelona Supercomputing Center}\\
Barcelona, Spain \\
jaume.abella@bsc.es}
\and
\IEEEauthorblockN{Francisco J. Cazorla Almeida}
\IEEEauthorblockA{\textit{High-Performance Embedded Systems} \\
\textit{Barcelona Supercomputing Center}\\
Barcelona, Spain \\
francisco.cazorla@bsc.es}
}

\maketitle
\thispagestyle{plain}
\pagestyle{plain}

\begin{abstract}
Multicore timing interference, arising when multiple requests contend for the same shared hardware resources, is a primary concern for timing verification and validation of time-critical applications. Bandwidth control and regulation approaches have been proposed in the literature as an effective method to monitor and limit the impact of timing interference at run time. 
These approaches seek for fine-grained control of the bandwidth consumption (at the microsecond level) to meet stringent timing requirements on embedded critical systems. 
This is typically achieved by resorting to ad-hoc, fine-tuned hardware and software configurations, exploiting hardware support, for monitoring bandwidth usage, and ad-hoc operating system level support, for inhibiting the execution of applications exceeding their allocated bandwidth. 
Such granularity and configurations, while effective, can become an entry barrier for the application of bandwidth control to a wide class of productized, modular ROS2 applications. This is so because those applications have less stringent timing requirements but would still benefit from bandwidth regulation, though under less restrictive, and therefore more portable, granularity and configurations.
In this work, we provide \textit{ROSGuard}, a highly-portable, modular implementation of a timing interference monitoring and control mechanism that builds on the abstractions available on top of a generic and portable Linux-based software stack with the Robotic Operating System 2 (ROS2) layer, a widespreadedly adopted middleware for a wide class of industrial applications, far beyond the robotic domain.
We deploy \textit{ROSGuard} on an NVIDIA AGX Orin platform as a representative target for functionally rich distributed AI-based applications and a set of synthetic and real-world benchmarks. We apply an effective bandwidth regulation scheme on ROS2-based applications and achieve comparable effectiveness to specialized, finer-grained state-of-the-art solutions.
\end{abstract}

\section{Introduction}
Complex heterogeneous \ac{mpsoc} \ac{cots} platforms are increasingly adopted, even in embedded critical domains, to support functionally-rich applications, often deploying machine learning based methods~\cite{Perez20}. 
In the automotive domain, for example, the trend towards Software-Defined Vehicles~\cite{liu2022sdv} and the market push towards autonomous driving and advanced driver assistance systems are clearly driving the integration of more and more complex software functionalities on the same platform. 
Such functionalities are delivered by data-intensive software applications building on computationally intensive models, like convolutional and deep neural networks. Those functionalities typically react to \ac{rt} stimuli from the environment they operate on, through an increasingly number of advanced sensors (e.g., cameras, lidar units, etc.) and actuators.
The required performance can only be met with complex \acp{hpec} 
platforms equipped with a large number of cores, \acp{gpu}, and application-specific accelerators
~\cite{Perez20}.

The adoption of \ac{hpec} platforms, however, seriously challenges timing predictability and analyzability, and complicates an already onerous verification and validation process~\cite{ManyCoresProblems}. In particular, the huge amount of shared hardware resources, including but not limited to shared memories and cache hierarchies, expose co-running applications to contention when attempting to simultaneously access the same shared hardware resources~\cite{HassanP20}. The impact of contention on an application execution time is known as \textit{multicore timing interference} and, if uncontrolled, can easily lead critical functions to miss their deadline~\cite{Interference2019} and, hence, impair the system's nominal behavior.

In the last decades, a lot of effort has been devoted, in industry and academia, to devise methodologies and tools to cope with multi-core timing interference \cite{yun2013memguard, yun2016memguard, izhbirdeev2024memcore, zuepke2023mempol}. Some solutions attempt to remove sources of interference in the platform altogether by enforcing full segregation across software partitions by leveraging hardware or software support fo resource partitioning. These solutions, however, despite effective in reducing interference in the system, are not able to completely prevent it, and often cause non-negligible performance degradation\cite{oliveira2025, costa2025}. More recently, a family of approaches has been proposed that aim at monitoring and controlling the use of shared hardware resources at run time, in view to protect critical applications from interference exceeding a predetermined threshold. These type of approaches, going under the umbrella term of \textit{memory bandwidth regulation} methods, provide a good compromise between interference control and overall performance \cite{izhbirdeev2024memcore, BandWatch, zuepke2023mempol, saeed2023}.

These approaches rely on hardware~\cite{MPAM23} or hypervisor/operating system-level~\cite{BAM,IsolationHyperV} support to implement fine-grained bandwidth usage monitoring and promptly react in case of over utilization. The monitoring aspect is dependent on the frequency at which it is possible to check for the bandwidth usage: the hardware and software support necessary to provide fine-grained interference control (below the $\mu s$ level) are not always available in \ac{hpec} hardware and software stacks and therefore cannot always be deployed without requiring important modifications to the setup. 

In fact, given their nature, achieving high levels of monitoring granularity may not be required for a wide class of systems. Availability and time-to-market considerations are pushing towards the adoption of non-specialized \ac{cots} platforms and Linux-based setups, often relying on the use of generic communication middlewares such as \ac{ros2}~\cite{ros2} to facilitate the design and implementation of large modular applications. For such applications, the level of control required for bandwidth regulation is typically looser than that considered by existing approaches, and contention control at the $ms$ level granularity is normally acceptable and still achieves an effective overall bandwidth control. Furthermore, it can help to avoid the risk of negatively affecting the overall system throughput~\cite{brilliWCET24}.

In this context, we propose \textit{ROSGuard}, a portable and modular bandwidth regulation solution that builds on top of the \ac{ros2} ecosystem, without requiring specialized hardware or software support. 
By reducing entry-access hardware and software requirements,
\textit{ROSGuard} extends the benefits of memory regulation to the wide class of \ac{ros2}-based productized software solutions that are not intended to be executed on top of specialized execution environments but rely on a generic Linux-based setup and naturally come with looser granularity requirements.
\textit{ROSGuard} builds on a \ac{ros2} environment and \textit{pthread} support to enforce thread priorities. It provides a fully functional and configurable bandwidth regulation solution that matches the granularity requirements of this type of applications. Moreover, \ac{ros2} offers comparable guarantees to those obtained by state-of-the-art methods on more controlled setups. 
Our key contributions are:
\begin{enumerate}
    \item We provide a highly-portable, modular, \ac{ros2}-based implementation of a memory bandwidth regulation mechanism, that better fits the requirements of productized, modular applications on \ac{ros2}. We discuss for the first time the main considerations when mapping bandwidth regulation concepts to the \ac{ros2} environment.
    \item As a proof-of-concept, we implement and deploy our mechanism on an NVIDIA AGX Orin target, representative of heterogeneous MPSoC in embedded domains.
    \item We demonstrate that our middleware-based mechanism offers comparable performance to that of more specialized methods, proving our approach as an useful alternative that enhances portability and facilitates the \textit{safe} deployment of large modular applications built over generic communication middlewares.
\end{enumerate}

The rest of the paper is structured as follows. Section \ref{sec:background} introduces the background concepts on \ac{ros2} and related work on memory bandwidth regulation techniques. Section \ref{sec:design} details the design and implementation of our proposal on \ac{ros2}. Section \ref{sec:evaluation} presents an evaluation of our approach and Section~\ref{sec:conclusion} concludes the paper.
\section{Background and Related Work}\label{sec:background}
In the following we provide the necessary background and related works on ROS2, especially form the stand point of \ac{rt} systems, and on  bandwidth regulation schemes.

\subsection{\ac{ros2} Basic Concepts and its Use in \ac{rt} Systems}
\ac{ros2} is the de-facto standard for modular, communication-based applications, with use-cases well beyond the original robotics domain. \ac{ros2} evolved from ROS, with the main purpose of improving predictability and providing \ac{rt} capabilities. 

In \ac{ros2}~\cite{ros2Journal}, a \textit{node} is the fundamental execution element or component that implements a specific software function. Nodes communicate with each other through \textit{topics}, according to publisher-subscriber semantics, to provide system-level functionalities. 
While more complex semantics for node interaction can be defined through \textit{actions} and \textit{services}, direct exchange of \textit{messages} through \textit{topics} is the baseline means for communication among nodes. 
The specific actions performed by a node in reaction to specific events (usually timers or topic updates) can be specified through a set of \textit{Callbacks}.

\begin{figure}[t!]
    \centerline{\includegraphics[width=1\linewidth, keepaspectratio]{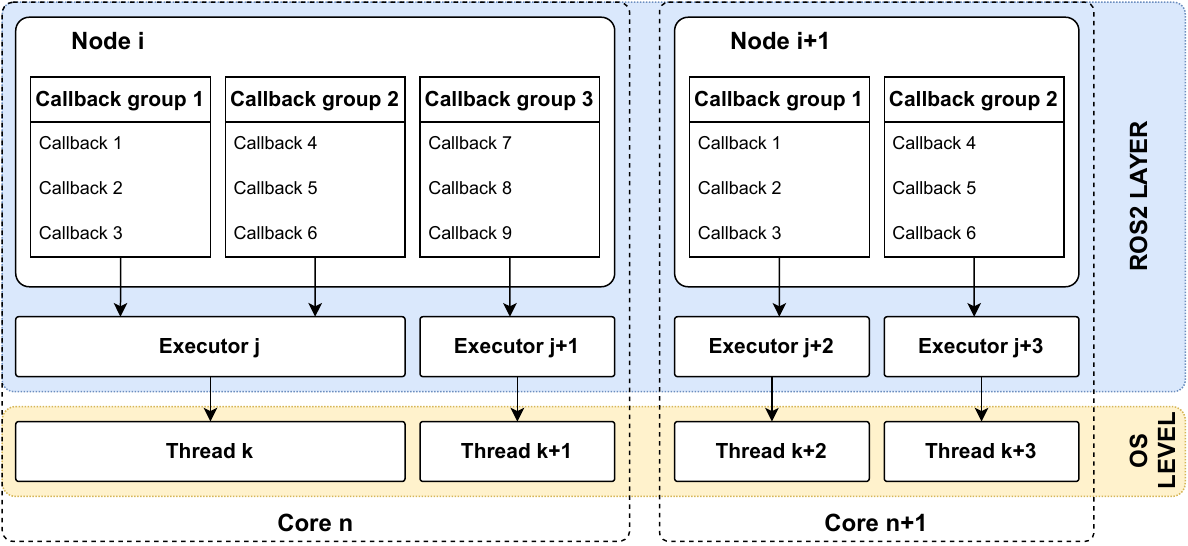}}\vspace{-0.2cm}
    \caption{Relation among \ac{ros2} elements and OS-level threads, exploited in \textit{ROSGuard} design and implementation.}
    \label{fig:ROS_elements}\vspace{-0.2cm}
\end{figure}

The \ac{ros2} layer relies on the concept of \textit{Executor} to identify the run-time entity that is responsible for executing the callbacks. Executors are typically mapped to a (set of) thread in the underlying operating system. While default Executors are assigned to nodes, it is still possible to instantiate a specific type of Executor (\textit{Multi-Threaded}, \textit{Single-Threaded}, and \textit{Static Single-Threaded}) to execute the callbacks of a node, depending on the callback semantics, (\textit{Mutally Exclusive} or \textit{Reentrant}), and intended behavior.
Callback groups, finally, offer a finer-grained control over the execution of callbacks as they allow to associate different callbacks to different executors, hence providing specific semantics within the same node~\cite{ros2Journal}. 
The relation among \ac{ros2} nodes, callbacks (and groups), executors, and OS threads is summarized in Figure~\ref{fig:ROS_elements}, where arrows represent the existence of a configurable mapping between the involved elements. In particular, it is possible to override \ac{ros2} internal scheduler by explicitly creating and associating threads at operating system level, during initialization, and associate them to specific executors. This, in combination with \textit{pthread} support for assigning thread priority at the OS-level, is at the basis of the \textit{ROSGuard} regulation mechanism.

\ac{ros2} has been increasingly deployed to realize complex modular functionalities in real-time embedded applications \cite{zhang2024, patel2023, abdulrahman2024}. However, there is still a gap between true \ac{rt} operation and the current performance of \ac{ros2}, mainly due to three reasons. First, the scheduling policies inside \ac{ros2} do not stick to a deadline-based strategy, jeopardizing \ac{wcet} analysis. Second, \ac{ros2} uses a non-deterministic dynamic memory allocation scheme, which can lead to delays, for instance, when a page-fault occurs. And third, \ac{ros2} is usually run over the Linux kernel, which neither does provide \ac{rt} performance out of the box.

There is a variety of works in the literature that evaluate the \ac{rt} performance of \ac{ros2} and aim to tackle such issues. 
Arafat et al~\cite{arafat2022} propose a deadline-based scheduling strategy for the \ac{ros2} executor and analyze the end-to-end response-times.
Krounaer et al.~\cite{kronauer2021} analyze the end-to-end latency of multi-node \ac{ros2} systems. 
Puck et al. \cite{puck2021} perform an in-depth evaluation of the \ac{ros2} communication-stack, and assess its current limitations. 
Choi et al \cite{choi2021} propose a priority-driven scheduler that prioritizes callbacks based on the given timing requirements of the corresponding chains, and they achieve predictable bounds on the end-to-end latency of critical chains. 
Dehnav et al.\cite{dehnavi2021} address the issue by implementing a hardware-software architecture (\textit{CompROS}) in a PS-PL platform. 
Casini et al \cite{casini2019} provide a practical analysis method to bound the worst-case response-times of \ac{ros2} applications. 
Delgado et al.~\cite{delgado2019} propose an architecture that supports priority-based scheduling of multiple tasks with results that indicate the proposed architecture can effectively provide an RT environment. As \ac{ros2} is run over Linux in most cases, other works \cite{gutierrez2018} evaluate the Linux communication stack for using it with \ac{ros2} in \ac{rt} applications, concluding that, with a proper configuration (including \ac{rt} kernel patch), ROS2 on Linux can reasonably meet \ac{rt} constraints.
More recently, the problem of guaranteeing scheduling properties across ROS2 and OS layers is tackled in~\cite{crost25}. We cope with these aspects by enforcing a controlled and exclusive mapping of callbacks, executors, and OS threads, as will be detailed later. 

The increasing adoption of \ac{ros2} to deploy applications with \ac{rt} requirements justifies massive research efforts on methodologies to ensure determinism and timing guarantees in less constrained execution platforms consisting of \ac{ros2} on top of mainstream Linux-based systems. \textit{ROSGuard} fits in this context by contributing a reference \ac{ros2} implementation for a timing-interference protection mechanism, with a regulation granularity and accuracy that is commensurate to the applications deployed into those systems, while providing an overhead comparable to that achieved with more specialized approaches,
but with an unmatched level of portability.

\subsection{Approaches to Memory Bandwidth Regulation}
Multicore timing interference in \acp{mpsoc} is a widely acknowledged concern in the design and verification of (mixed-criticality) time critical embedded systems, where \textit{freedom from interference}~\cite{ISO26262} is a prerequisite for system qualification. 
Contention arising from massive sharing of hardware resources causes highly variable access latencies, with potentially overwhelming cumulative impact on execution time and ultimately jeopardizing the derivation of trustworthy timing guarantees on software.
Full temporal and spatial isolation across co-running applications is not feasible to achieve without severely compromising performance~\cite{Perez20}. For the type of systems addressed in this paper, overall performance plays a critical role, possibly as important as timely execution.

This aspect led industrial and academic researchers to focus on more efficient methods based on bandwidth regulation for limiting the sources of timing interference while at the same time exploiting overall performance. The general mechanism of these approaches consists in \textit{sampling} the bandwidth utilization of non-critical applications or tasks and triggering a \textit{regulation} mechanism (e.g., throttling or core disabling) whenever they exceed a predefined threshold, deemed acceptable for co-running critical applications. Bandwidth regulation methods do not exclude, and rather can complement, existing \ac{qos} and hardware partitioning support (e.g., cache and memory partitioning).

Several approaches have been proposed in the literature to implement memory bandwidth regulation schemes, exploiting hardware and software level mechanisms. Bandwidth regulation approaches using ad-hoc or platform-specific hardware modules~\cite{farshchi2020,MCCU,BandWatch} can guarantee fine-grained regulation with cutting-edge performance. However, they rely on ad-hoc hardware-level support that is not necessarily available in every \ac{hpec} \ac{cots} target.

Another family of approaches, instead, enforce bandwidth regulation mainly via software-level mechanisms~\cite{yun2013memguard, yun2016memguard}. Software methods, when compared to purely hardware ones, usually support looser monitoring granularity in exchange of more flexibility as they only partially build on (less-specialized) hardware support~\cite{izhbirdeev2024memcore,zini,Alejandro21}.
All these approaches build on the capability to monitor the hardware resource-usage via the \ac{pmu}, counting hardware events at different scopes (e.g. core or memory controller).

On this line, \textit{MemGuard} \cite{yun2013memguard, yun2016memguard} is one of the earliest and maybe the most common variant of regulation methods building on core-level \ac{pmc} sampling. The maximum allowed bandwidth for non-critical applications is based on the worst-case (guaranteed) bandwidth, hence the total usage of the real available bandwidth is rather conservative. Also, it relies on interrupts both for the sampling and core-regulation, which causes considerable execution-time overheads. This work has been followed by various other contributions \cite{martins2020, modica2018, dagieu2016}.

Periodic \ac{pmc} polling has been shown to achieve good results. Consequently, there is a bunch of works that use this approach. \textit{MemPol} \cite{zuepke2023mempol}, for instance, distributes available memory bandwidth to each core while taking global bandwidth into consideration. It also moves the execution of the controlling logic out of the processor cores to other elements in the \ac{soc} in order to reduce execution-time overhead. As various hardware events are available in a modern \acp{pmu}, complex regulation strategies can be designed, such as the one in \cite{saeed2023}, which proposes a probabilistic distribution-driven memory regulation.

Some modern platforms expose interfaces that allow monitoring memory-usage at the \ac{dram} controller, which enable precise measurement of memory-bandwidth. Some authors \cite{saeed2022dr, sohal2020ewarp} leverage this feature to design regulation mechanisms, usually setting those \acp{pmc} as countdown counters that trigger interrupts when emptied. This causes the available regulation policies to be limited as typically various \ac{pmc} values cannot be combined. Moreover, performance counters at the \ac{dram} controller are not always available.

With respect to the regulation mechanism, several methods build on the OS support to enforce core throttling in the non-critical tasks~\cite{yun2013memguard}. \textit{MemCore}~\cite{izhbirdeev2024memcore}, instead, is a hardware-assisted mechanism that uses ARM \textit{CoreSight}, the debug interface available in ARM platforms, to halt the offending cores, achieving superior performance and nanosecond-scale regulation. However, it is targeted to \acp{soc} with integrated FPGAs, so it is not widely applicable. {\color{white}\ac{nrt}}

State-of-the-art approaches provide a wide spectrum of solutions that can effectively control the impact of contention with variable sampling and regulation granularities, ultimately offering different trade-offs between guarantees and performance overheads. \textit{ROSGuard} builds on the same concepts contributed by state-of-the-art bandwidth regulation approaches to offer a generalizable solution that extends the benefit of bandwidth regulation methods to a wider class of productized systems 
built on \ac{ros2} that rely on less restrictive and controlled setups than those typically deployed for the most critical functions in real-time embedded systems. 
\textit{ROSGuard} provides a highly-portable, modular, and configurable solution that only requires a \ac{ros2} environment on the software side, and minimal support for \ac{pmc} sampling on the hardware side. \ac{ros2} is a widely adopted middleware framework for productized applications used across industrial domains that runs on virtually any system running a Linux distro, and memory-mapped \acp{pmc} are abundant in current embedded targets.
\section{ROS2-based Bandwidth Regulation Scheme}\label{sec:design}
\textit{ROSGuard} deploys a fully-functional bandwidth regulation solution on top of the \ac{ros2} middleware, with enhanced portability and modularity. The overall mechanism is consistent with state-of-the-art bandwidth regulation approaches (e.g.,~\cite{zuepke2023mempol}) and aims at limiting the memory bandwidth used by \textit{regulated (non-critical) components} below a predefined threshold, so that the contention delay possibly suffered by critical components on the same target is minimized. To this extent, a \textit{controller component} is responsible for keeping track of the bandwidth (cumulatively or singularly) used by a regulated one, often by gathering information from hardware events through the \ac{pmu}. In case the allocated bandwidth is exceeded, the controller temporarily inhibits the execution of regulated components up to the next regulation window, where bandwidth budget will be replenished. Effectiveness and overhead of a regulation mechanism depend on the frequency at which hardware events are gathered (\textit{sampling}).

Bandwidth regulation approaches are typically enforced at core-level \cite{yun2013memguard,zuepke2023mempol,saeed2023,sohal2020ewarp} where mixed-criticality tasks are allocated to distinct sets of cores depending on the criticality level or real-time requirements, to exploit architectural level segregation, and favor partitioning of resources. Regulation mechanism can be adapted to operate at task level, on a finer grained regulation scope, but cores generally provide an appropriate granularity of control. \textit{ROSGuard} differentiates from existing approaches in that it is fully defined at the ROS2 middelware level: the controller, regulated and critical elements are all \ac{ros2} nodes. Furthermore, the sampling and regulation logic are completely realized on top of \ac{ros2} functionalities. 
While the mechanism can be applied at different granularity-levels (task, core, component), we consider each ROS2-level component statically mapped to a core, building on available support for affinities. Accordingly, applications or nodes allocated to specific cores, are referenced to as either \ac{rt} (critical) or \acs{nrt} (non-RT a.k.a. best effort or non-critical), depending on the nature of the applications themselves.

\begin{figure}[tbp]
    \centerline{\includegraphics[width=1\linewidth, keepaspectratio]{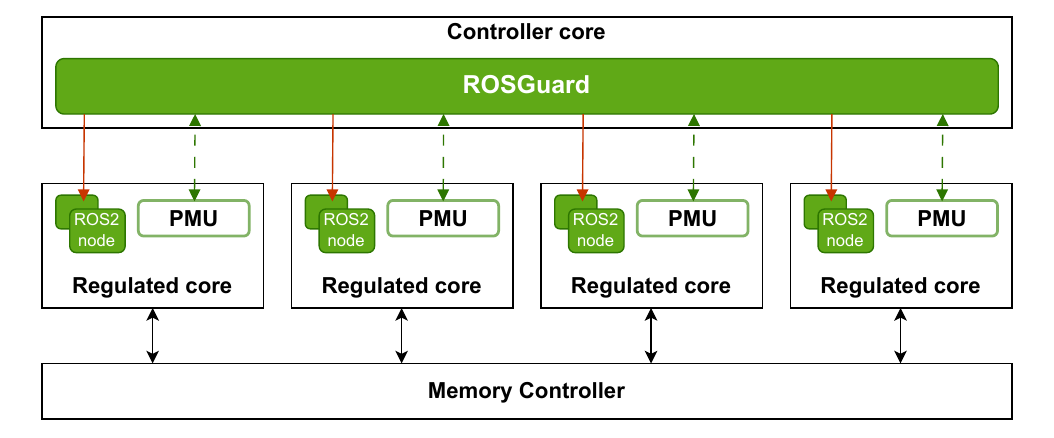}}\vspace{-0.2cm}
    \caption{Overview of the \textit{ROSGuard} components. The red arrows indicate control-actions. The green arrows indicate the feedback-loop.}\vspace{-0.2cm}
    \label{fig:overview}
\end{figure}

\subsection{\textit{ROSGuard} Main Components and Flexibility}
Figure~\ref{fig:overview} summarizes the interaction between regulated and controller nodes in \textit{ROSGuard}. The controller \ac{ros2} node (ROSGuard node in the picture) is deployed on a dedicated core and keeps track of the bandwidth consumption of regulated cores, which in turn can execute one or multiple \ac{nrt} \ac{ros2} nodes. 
The controller node, whenever the bandwidth allocated to \ac{nrt} nodes is exceeded, can halt these same nodes, forcing them into a throttle state.

A peculiarity of \textit{ROSGuard} is that it builds on a flexible design that supports the deployment of different \textbf{sampling} schemes and \textbf{regulation logic} on top of the same software architecture model, with minimal changes on the semantics of the \textit{ROSGuard} software component archetypes. 
\begin{itemize}
    \item \textit{ROSGuard} supports \textit{self-sampling}, where \ac{nrt} nodes periodically send updated \ac{pmu} values to the controller node.
    \item \textit{ROSGuard} also supports \textit{external sampling}, where the controller itself directly polls the regulated component from the outside. 

\end{itemize}    
It is also possible to adjust the \textbf{scope} of the considered hardware events, e.g. to single nodes or all nodes in a core. 

\textit{ROSGuard} regulation logic can be applied on configurable \textbf{timing intervals} so that inhibition of \ac{nrt} nodes is enforced within the interval boundaries rather than the completion of the \ac{rt} application.
    
Finally, on the protection logic, it is possible to define specific\textbf{ regulation conditions} by carefully defining the regulation window and selecting the hardware events to be considered in the computation of the bandwidth usage.  

\textit{ROSGuard} builds on ROS2 support for \textit{Callback\_groups} and \textit{Executors}~\cite{ros2Journal} and generic \textit{pthread} library support. These features allow fine grain control of how nodes are executed in the OS layer by enabling the definition of multiple executors with different priorities to be mapped to the same core. While, by default there is no correspondence between \ac{ros2} executors and OS-level threads, it is possible to exploit pthread library support to explicitly map executors to OS-level threads and specific callbacks. In practice, by combining callback, executors, and thread priorities, it is possible to define a priority-driven hierarchy of callbacks that can always pre-empt each other according to specific semantics. If we encode the standard (low-priority) thread responsible for the callback corresponding to the node nominal behavior, then a high-priority thread can be associated to specific callbacks to perform critical and urgent actions for the bandwidth regulation mechanism, as sampling the \ac{pmc} values or inhibiting the execution of non-critical nodes.

\subsection{\textit{ROSGuard} Addressed Challenges and High-Level Design}
Moving from task/core level to middleware components incurs specific challenges.
On the one hand, some aspects in the bandwidth regulation mechanism are straightforwardly achieved on top of the publisher-subscriber semantics: the exchange of sampling information between the \ac{nrt} nodes and the controller, for example, can be implemented as a topic.
On the other hand, bandwidth regulation requires means to control the execution of \ac{nrt} nodes, with the capability to temporally inhibit their execution, which requires to ensure a direct correspondence between \ac{ros2} nodes and run-time entities as processes and threads on the underlying OS. 

The inhibition mechanism in \textit{ROSGuard} is obtained by combining \ac{ros2} concepts of \textit{Callback\_groups} and \textit{Executors}~\cite{ros2Journal} to define multiple callbacks for the same nodes and map them exclusively to different executors. Each callback is then exclusively associated to a dedicated executor. 
These executors, in turn, can be mapped to a set of OS-level threads running on the same core but with different priorities, leveraging the \textit{pthread} library support. 
Ultimately, this allows to define a priority-driven hierarchy of (preemptible) callbacks according to the bandwidth regulation semantics. In particular, for a given \ac{nrt} node, we can define:
\begin{itemize}
    \item A low-priority thread that is responsible for the callback corresponding to the nominal behavior. 
    \item A (set of) high-priority thread is associated to specific callbacks to perform critical and urgent actions for the bandwidth regulation mechanism, and specifically inhibiting the execution of the low-priority thread.
    \item Optionally, an intermediate-priority thread is responsible for sampling the \ac{pmu} values, if \ac{nrt} self-sampling mode is implemented. 
\end{itemize}

Before entering into \textit{ROSGuard} implementation details, we discuss its main design-level characteristics from the perspective of state-of-the-art bandwidth regulation mechanisms.  

\subsubsection{\textbf{Sampling method and metric}} The considered sampling method and metric determine how the activity of nRT applications is monitored. Modern microprocessor architectures (e.g., ARMv8-A) feature rich \acp{pmu} that allow monitoring a variety of hardware events on different components (e.g., core, memory controller) at different granularity levels (e.g., thread, process, core). These hardware events can be combined to derive resource usage metrics and accurate estimates of bandwidth usage for different components. For this task, \textit{ROSGuard} just builds on the generic support provided by the \texttt{perf-tool} layer (available in all Linux distributions).
\textit{ROSGuard} default behavior exploits last-level cache (LLC) events to measure memory bandwidth usage as regulation metric. However, it is easily configurable to sample a configurable set of events and define a custom regulation metric.

Also, \textit{ROSGuard} can support both self- and external-sampling: in the latter case, sampling events are polled by the bandwidth regulation module at a fixed interval; in the former, instead, events are provided by the the core itself possibly with a different frequency than that of the control mechanism.
In any case, a careful selection of the sampling interval offers a trade-off between the sensitivity of the control (i.e., how fast it is able to detect bandwidth exceedance) and the performance overhead. Both are explored, under different configurations, in Sections \ref{sec:pmcimpact}, \ref{sec:reaction}.
Possible overshoots in the bandwidth allowed for \ac{nrt} applications can be compensated by setting a threshold-setpoint slightly lower than the desired maximum allowed bandwidth. The bandwidth tolerance threshold will affect the performance of the mechanism in terms of precision and execution-time overhead. We assess the impact of those parameters in Section \ref{sec:evaluation}. 

\subsubsection{\textbf{Regulation Logic}}
Regulation schemes can differ on the granularity at which bandwidth control is enforced.
Depending on the setup, regulation logic can be applied synchronously or asynchronously with respect to sampling.
When the regulation logic is applied, the metric considered for the regulation is assessed against the predefined setpoint. If such value is exceeded, the control module signals the corresponding (\textit{non-critical}) cores to stop consuming further bandwidth. We evaluate the effect of setpoint selection in Section \ref{sec:threshold}.

\textit{ROSGuard} regulation logic can be applied on configurable timing intervals so that inhibition of \ac{nrt} nodes is enforced within the interval boundaries rather than the completion of the \ac{rt} application. 
While inhibiting cores until completion of \ac{rt} applications offers best execution-time guarantees, it also causes considerable slowdowns in the execution of \ac{nrt} applications, which limits the overall throughput and might not be acceptable in general. 
We evaluate different regulation intervals in Section \ref{sec:evaluation}. We show how appropriate granularity trade-offs yield quite good results, ensuring adequate protection while also preserving \ac{nrt} applications' performance.

\subsubsection{\textbf{Inhibition mechanism}}
The inhibition mechanism ensures that non-critical tasks that have exceeded their maximum bandwidth usage are prevented from causing further interference on \ac{rt} tasks. While hardware-specific support exists \cite{izhbirdeev2024memcore}, by using a software-level mechanism, our solution guarantees maximum portability. We enforce the cores throttling by triggering the callback associated to a higher-priority thread, which will preempt the non-critical task (i.e. executing the nominal callback) and simply enter into an empty idle loop. While such higher-priority task is executing, no request will be generated by those cores on any shared hardware resource, thus ensuring exclusive usage to the \ac{rt} application. As will be detailed next, \textit{ROSGuard} exploits the definition of \texttt{callback\_groups} and multiple \texttt{executors} with different priorities to achieve the effect of core throttling.

\begin{figure}[tbp]
    \vspace{-0.2cm}\centerline{\includegraphics[width=0.9\linewidth, keepaspectratio]{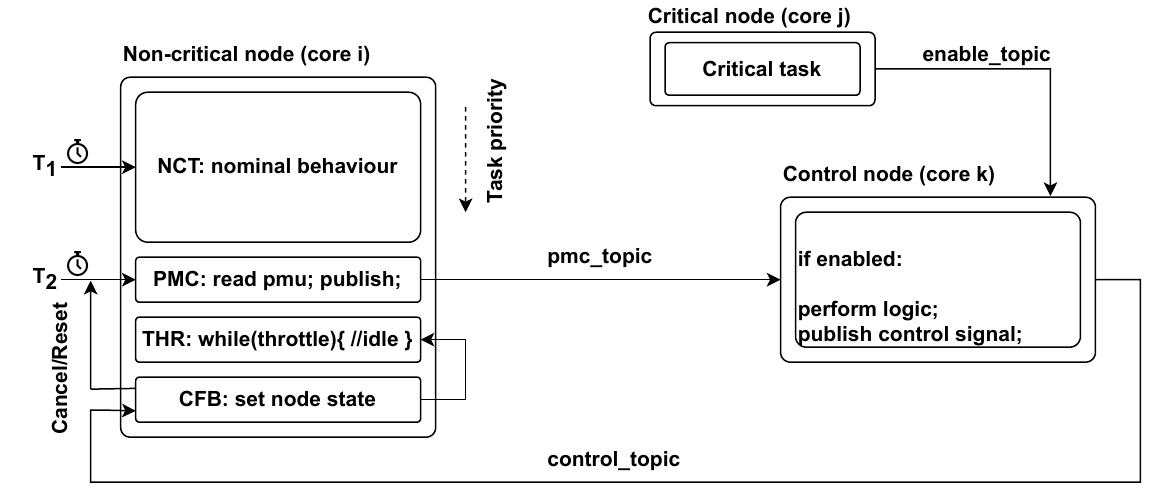}}\vspace{-0.2cm}
    \caption{Minimal \textit{ROSGuard} setup.}
    \label{fig:ros_nodes}\vspace{-0.2cm}
\end{figure}

\subsection{\textit{ROSGuard} Archetypes and Implementation}
The \textit{ROSGuard} mechanism builds on specialization of \ac{ros2} nodes (archetypes) that can be deployed in a modular way to realize a more or less wide and complex regulation scheme. 
In its minimal setup, \textit{ROSGuard} comprises three components, corresponding to node archetypes: an \ac{rt} node, an \ac{nrt} node, and a controller node, as shown in Figure~\ref{fig:ros_nodes}. In our minimal setup, nodes are deployed to distinct cores but the same core can be shared, e.g. by tasks of the same criticality.

The bandwidth regulation mechanism builds on the capability of enforcing different \textit{execution modes} on the nCT, depending on the bandwidth status. The nodes cooperate through a set of ROS2 topics that are responsible for sharing the sampled \ac{pmu} values and for controlling the execution mode. 
The basic functionality of those nodes is the following:
\begin{itemize}
    \item The control node is the \ac{ros2} node enabling the monitoring interval, setting the execution mode of the non-critical nodes based on the control logic and the associated metrics, and eventually replenishing the bandwidth budget at the beginning of each monitoring interval. 
    The control node is enabled upon each activation of the critical node and, in turn, activates the monitoring interval.

    \item The critical node executes an \ac{rt} application either periodically or in reaction to an external event, depending on the application. The activation pattern is irrelevant for the regulation mechanism. The node will send an enable signal at the beginning of the \ac{rt} application to the control module and a disable signal at application completion, as the bandwidth monitoring and regulation is only necessary during the execution of the \ac{rt} applications.
    
    \item The non-critical node executes an \ac{nrt} (non-critical or best-effort) application. Under a self-sampling scheme 
    the node periodically publishes \ac{pmc} values to the topic at which the control node is subscribed. Alternatively, the control node can sample \ac{pmc} values directly, from an external core. The node supports three execution modes, corresponding to three internal states, triggered by the so-called \textit{control topic}: when \textit{OFF}, it will run the \ac{nrt} application normally (nominal behavior); when \textit{ON} it will run the \ac{nrt} application but also monitor the \ac{pmu} and publish the collected measurements to the control node. When in \textit{THR} (\textit{throttle}) state, it will idle until another state-change signal is received through the control topic.
    
\end{itemize}

The implementation of the non-critical node deserves some more details as it heavily builds on fine-grained control of executors, callbacks, and thread priorities. ROS2 capabilities to assign each callback exclusively to a specific thread in the underlying OS, in combination with the definition of thread priorities and core affinities, gives the programmer full control over the execution priorities of the \ac{ros2} callbacks. Under \texttt{SCHED\_FIFO} priority-driven preemptive scheduler, a careful priority assignment to executors allows to enforce and exploit preemption among callbacks mapped to the same core.
We exploit such callback-preemption feature as the mechanism to enforce state transitions within a non-critical node where each status enables only specific callbacks. 

\begin{figure}[tbp]
    \centerline{\includegraphics[width=0.8\linewidth, keepaspectratio]{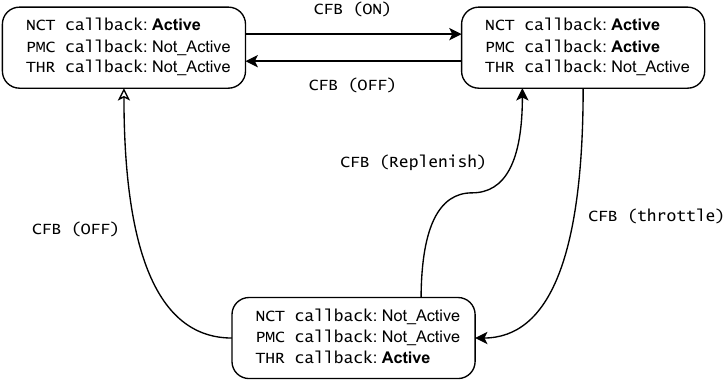}}\vspace{-0.2cm}
    \caption{State transitions diagram in \textit{ROSGuard} design.}
    \label{fig:state_diagram}
\end{figure}

The non-critical node implements four callbacks $CFB$, $NCT$, $THR$, and $PMC$ as shown in Figure~\ref{fig:ros_nodes}. 
A highest-priority callback ($CFB$) controls the transitions among states in the node.
It can always preempt other running callbacks to react to the commands received from the control node and sets the state of the node accordingly into \textit{nominal}, \textit{sampling}, and \textit{throttling} states.
In the nominal state, the node \textit{nominal} behavior is implemented as a lowest-priority callback ($NCT$).
If self-sampling is enabled, the sampling state is entered by sending a request to enter into the \textit{ON} state: in this case, the lowest-priority callback $NCT$ will still run but will be periodically preempted by the $PMC$ callback sampling the \ac{pmu} and publishing the collected values. 
If a transition to the \textit{throttling} state is requested, a throttle command triggers a second-highest priority callback ($THR$) that executes an infinite idle loop until a different command is received by the $CFB$. 
If an \textit{OFF} command is received, the node will move back to the nominal state, either from the sampling or throttling states, and the $NCT$ callback will resume its execution (and self-sampling will be disabled). 
Finally, a \textit{replenish} command, issued at the end of each control interval, will trigger the transition from the throttling state to the sampling one.
Note that the $THR$ callback is preventing $NCT$ and $PMC$ callbacks to execute, thus enforcing the throttling behavior.
In the case of external sampling, the $PMC$ callback is not implemented.
The state transitions and status of the relevant callbacks are summarized in Figure~\ref{fig:state_diagram}. 

In general, \textit{ROSGuard} can be deployed with any number of critical and non-critical nodes, limited only by the number of cores available in the platform or other performance limitations. This allows for a variety of configurations, making our approach suitable with many applications, platforms and requirement sets. For instance, a system may run all \ac{rt} applications in a single core, dedicate another core for control, and spread all other (\ac{nrt}) functionalities throughout the remaining cores.
The control node is run on a dedicated core to uncouple the regulation from the execution of the actual applications.

\section{Evaluation}\label{sec:evaluation}
We evaluate \textit{ROSGuard} effectiveness, in protecting \ac{rt} applications, and its incurred overhead on regulated applications. Both dimensions largely vary as a result of trade-offs between accuracy (control granularity) and overall system performance. For this reason, we assess \textit{ROSGuard} in its minimal setup with one node per archetype, under different parameter configurations for sampling period, regulation period, and bandwidth threshold. 
It is worth recalling that \textit{ROSGuard} is designed to provide bandwidth regulation capabilities on productized systems deployed on top of the \ac{ros2} middleware, hence with looser control granularity constraints than those sought for typical time-constrained embedded systems functionalities.
Nonetheless, when possible, obtained results are compared with results reported by state-of-the-art bandwidth regulation approaches on the same benchmarks. In particular, we focused on \textit{MemGuard}~\cite{yun2013memguard} as it represents a reference solution for bandwidth regulation, with a control-logic and mechanism comparable with those supported by \textit{ROSGuard} design.

\subsection{Experimental Setup}
\begin{figure}
    \centerline{\includegraphics[width=0.9\linewidth, keepaspectratio]{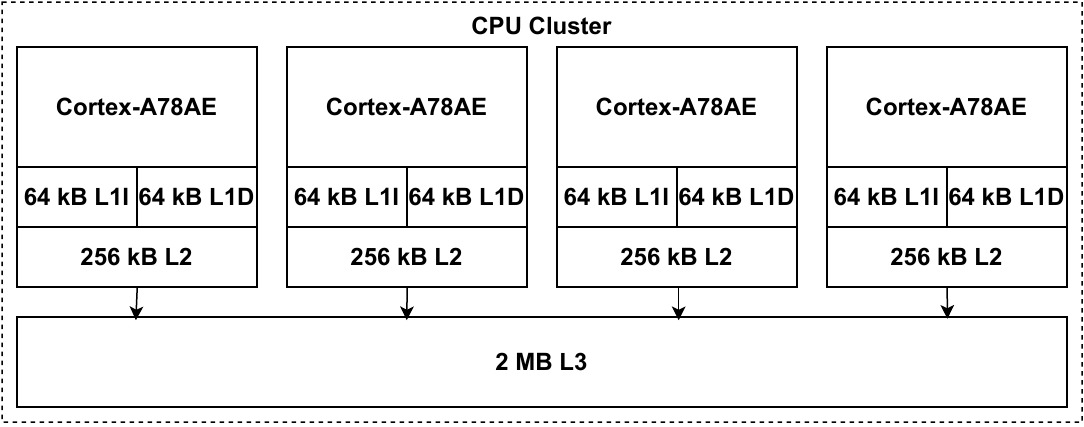}}\vspace{-0.2cm}
    \caption{Architecture of one CPU cluster in the NVIDIA Jetson AGX Orin.}
    \label{fig:orin_cluster}\vspace{-0.2cm}
\end{figure}

We evaluate ROSGuard on a representative setup for functionally-rich embedded applications including an NVIDIA Jetson AGX Orin running an adapted Ubuntu 22.04 (Linux-tegra 5.15) OS and \ac{ros2} (Humble version).
Focusing on the Core Complex, NVIDIA Jetson AGX Orin 64GB Development Kit~\cite{orinTRM} comprises three \ac{cpu} clusters, each one featuring four ARM Cortex-A78AE cores, with 64 kB L1 data and instruction caches, 256 kB L2 cache and a cluster-shared 2 MB L3 cache, as shown in Figure~\ref{fig:orin_cluster}. Following standard practice for time-constrained systems on Linux setups, we configured the system to minimize the OS-induced noise (e.g. RT-patch, kernel parameters for core isolation and irq mapping) and fixed the platform power model and maximum cores' frequency at $2201$ megahertz.

In consideration of the architectural clusterization of the platform and cache-hierarchy, 
in our experiments we focused on intercepting and controlling timing interference already at the level of the cluster-shared L3 cache
as it allows to minimize the propagation of interference across clusters. Accordingly, we configured \textit{ROSGuard} regulation metric to model bandwidth usage using L3-related \ac{pmu} events.

We consider bandwidth usage for L3 cache accesses per core ($l3\_accesses$) using \ac{pmu} events for L2 cache misses (write-backs and refills) as shown in Eq.~\ref{eq:bw}, where $64$ is the cache-line size in Bytes, $cycles$ stands for CPU cycles (from \ac{pmu}), $freq$ is CPU frequency in Hertz, and 1024*1024 is a conversion factor.
\begin{equation}\label{eq:bw}
\footnotesize
    bw = \frac{64*freq*l3\_accesses}{cycles*1024*1024}
    \newline
\end{equation}

However, it is worth recalling sampling and control logic in \textit{ROSGuard} are customizable, with different metrics and events.

As workloads for \ac{ros2} nodes, we rely on both synthetic and real-world applications, in line with related works. 
As a synthetic workload, we considered the Isolbench suite \cite{isolbench, valsan2016isolbench}, a set of particularly aggressive synthetic benchmarks designed to evaluate contention in shared resources. We run \textit{bandwidth\_read} and \textit{bandwidth\_write} with a working set size of $2048$ kilobytes.
As real-world workloads, we focused on the \ac{sdvbs} \cite{venkata2009sdvbs}, which is used in related works \cite{izhbirdeev2024memcore, zuepke2023mempol, saeed2022dr} and comprises 9 state-of-the-art computer-vision applications, tailored by the University of California in collaboration with vision researchers. The input datasets for the applications come in a variety of sizes. We use the \textit{vga} resolution datasets so the working sets fit in the L3 cache of the platform, in line with the evaluation objective. Average Instructions Per Cycles (IPC) and Bandwidth usage for the considered applications, when deployed on the target platform, are shown in Table \ref{tab:benchmarks}.

\begin{table}[tbp]
\centering
\caption{Comparison of benchmarks.}
\scriptsize
\begin{tabular}{|l|l|c|c|}
\hline
\textbf{Suite} & \textbf{Benchmark} & \textbf{Avg. IPC} & \textbf{Avg. bandwidth (MB/s)} \\ \hline
\multirow{2}{*}{Isolbench} & \textit{bandwidth\_read} & $1.07$ & $26271.74$ \\
              & \textit{bandwidth\_write}             & $1.61$ & $25519.15$ \\
\hline
\multirow{5}{*}{SD-VBS} & \textit{disparity}                     & $0.83$ & $1517.16$ \\
                       & \textit{mser}                          & $2.16$ & $3722.43$\\
                       & \textit{sift}                          & $0.58$ & $232.15$ \\
                       & \textit{stitch}                        & $0.69$ & $713.08$ \\
                       & \textit{tracking}                      & $3.89$ & $393.43$ \\
                       \hline
\end{tabular}
\end{table}\label{tab:benchmarks}

\subsection{Sampling Overhead}\label{sec:pmcimpact}
The first set of experiments aims to assess the impact of a sampling mechanism based on \ac{ros2} can cause on the execution-time of the \ac{nrt}. The overhead is composed of the cost of polling the \ac{pmu} in the \ac{nrt} node and, in case of self-sampling, publishing the obtained values for making them available to the control node. 

\begin{figure}
    \centering
    \includegraphics[width=1\linewidth, keepaspectratio]{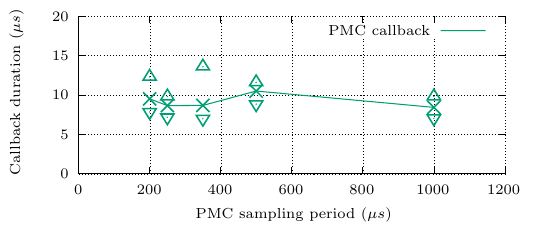}
    \vspace{-0.2cm}
    \caption{Distribution of the duration of the $PMC$ callback. The plot shows the minimum ($\triangledown$), maximum ($\triangle$), and median ($\times$) value for each sampling period.}
    \label{fig:single_pmc}
\end{figure}

We first measure the delay introduced by a single reading of the \acp{pmc} on the $PMC$ callback. We show the result for various sampling periods in Figure~\ref{fig:single_pmc}.
We observe a slight variation in minimum ($\triangledown$), maximum ($\triangle$), and median ($\times$) values.
The duration of the callback, however, can be typically considered in the order of $15\mu s$.

\begin{figure}
     \centering
    \includegraphics[width=1\linewidth, keepaspectratio]{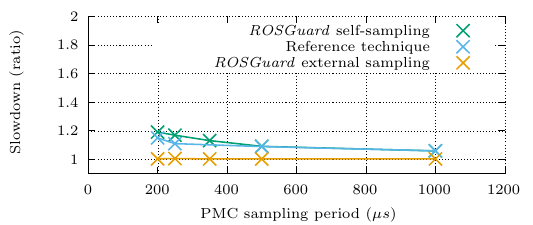}\vspace{-0.3cm}
    \caption{Impact of sampling the \ac{pmu} on the execution-time overhead of an application running in isolation. Comparison to a state-of-the-art reference implementation~\cite{yun2013memguard}.}
    \label{fig:pmc_impact}
    \vspace{-1.5em}
\end{figure}

Figure~\ref{fig:pmc_impact}, instead, shows the cumulative impact of the $PMC$ callback in the self-sampling scenario, where the \ac{nrt} node is responsible for making \ac{pmu} data available to the control node. 
We report the overhead suffered by an example application (\textit{bandwidth\_read}) when varying the \ac{pmu} sampling period, relative to its execution time in isolation. 
The relative slowdown incurred by (interrupt-based) self-sampling in MemGuard, as reported in ~\cite{yun2013memguard,zuepke2023mempol}, is also plotted as a reference.
A first observation from the results is that, as the cost of a single sampling callback is rather negligible, the cumulative sampling overhead is reasonably low and quite good for sampling periods from $500$ $\mu s$ and higher, in line with the granularity required for the types of application normally deployed on \ac{ros2}. 
Interrupt-based sampling implemented in \textit{MemGuard} shows comparable overheads.
While finer granularity, (lower than $100\mu s$) would lead to excessive overheads, that level of granularity is not required for the type of applications we are targeting.
Results on the cumulative overhead of self-sampling also allow to derive the approximate overhead incurred by the underlying \ac{ros2} layer. For instance, when running at a $1$ kilohertz sampling-rate, a slowdown-ratio of $1.059\%$ indicates an absolute increase of $59\mu s$ per $ms$ in execution-time, that means $59\mu s$ for each sampling period, since the \ac{pmu} polling is running at $1$ kilohertz. Considering that a single \ac{pmc} callback takes about $15\mu s$, the remaining time of approximately $44\mu s$ arguably comes from the \ac{ros2} communication scheme and underlying processes.

Finally, it is worth recalling that such overhead can be avoided if sampling is performed from outside the controlled node~\cite{zuepke2023mempol}. \textit{ROSGuard} can be also configured to sample the \ac{pmu} from outside the regulated core, by the control node itself, hence mimicking \textit{MemPol} \cite{zuepke2023mempol} behavior. Figure~\ref{fig:pmc_impact} also reports the impact of sampling on the regulated node when sampling is performed from outside (external sampling). In this case, we still observe a negligible impact induced by the interference arising when sampling the \ac{pmu}.

\subsection{Activation Delay}\label{sec:reaction}
In the following experiment, we considered the mechanism activation delay, i.e. the time it takes from the moment a \ac{pmc} sample incurring a threshold exceedance is published to when the regulation mechanism takes action (i.e. the throttle callback is triggered). 
This metric is critical for the effectiveness of the regulation mechanism as, in that interval, the \ac{nrt} node is still allowed to consume bandwidth.
Again, the end to end delay depends on the implementation of the sampling scheme.

Under the self-sampling paradigm (baseline implementation in \textit{ROSGuard}), such interval involves various steps in the node-chain responsible for the monitoring and regulation:

\begin{enumerate}
    \item The $PMC$ callback on the \ac{nrt} node must publish the sampled \ac{pmc} values.
    \item The control-node receives the message carrying the values and triggers the corresponding callback.
    \item The callback triggered on the control-node executes the control-logic.
    \item The control-node publishes the control-action. In this case, signals the \ac{nrt} node to change its state to \textit{THR} (\textit{throttle}).
    \item The \ac{nrt} node receives the message, which triggers the $CFB$ callback.
    \item Execution of the $CFB$ callback activates a \textit{zero-time} timer that \textit{instantaneously} triggers the throttle callback.
\end{enumerate} 

Instead, if sampling in performed externally, by the control node, only the last three steps must be completed.

\begin{figure}
     \centering
    \includegraphics[width=1\linewidth, keepaspectratio]{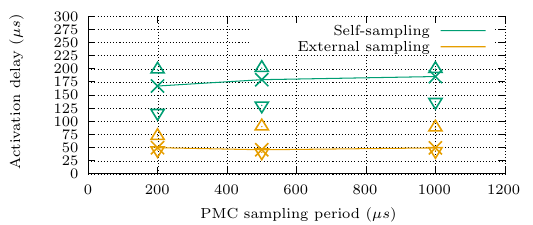}\vspace{-1em}
    \caption{Distribution of the activation delay, i.e. the time measured from publishing a threshold-exceeding \ac{pmc} value, to when the throttle callback is entered. The plot shows results for the minimum ($\triangledown$), maximum ($\triangle$), and median ($\times$) values, when using the self- and external-sampling techniques.}
    \label{fig:act_delay}
    \vspace{-1.5em}
\end{figure}

Figure~\ref{fig:act_delay} reports the activation delay when varying the sampling period. For the self-sampling scenario, as expected, the delay is almost constant, essentially unaffected by the sampling period, and stays below $200$ $\mu s$. A safety-margin on the application's bandwidth-threshold allows to compensate for such delay.
In terms of overhead, delay on activation is relatively negligible, compared to that caused by sampling, as the former only occurs when a bandwidth-budget exceedance occurs on a \ac{nrt} node.

In case of external sampling, as expected, we observe a considerable decrease in the activation-delay as the control node can immediately trigger the inhibition of the \ac{nrt} node as soon as it detects a threshold being exceeded. Also in this case, delay is almost constant across sampling-periods, always below $100$~$\mu s$, with median values very close to minima.

\subsection{Sensitivity Analysis: Setpoint Selection}\label{sec:threshold}
\begin{figure}[tbp]
     \centering
    \includegraphics[width=1\linewidth, keepaspectratio]{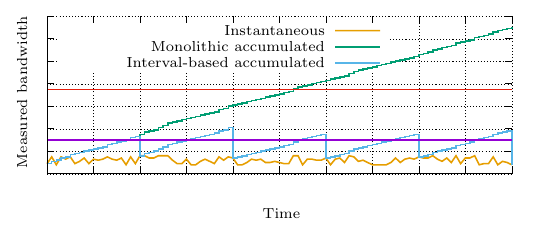}\vspace{-0.3cm}
    \caption{Theoretical bandwidth measured (orange) and accumulated (light blue) under interval-based regulation policy, compared to overall accumulated bandwidth (green) without considering regulation intervals. 
    The purple and red horizontal lines represent example bandwidth-threshold values for interval-based and monolithic regulation policies respectively.} 
    \label{fig:control_policies}
    \vspace{-1em}
\end{figure}

In the following experiments, we explore the effect that bandwidth-threshold \textit{setpoint} has on the sensitivity and performance of the regulation. We aim at gauging the impact of this metric on the execution of both \ac{rt} and \ac{nrt} nodes.
In the assessment, we also consider what we call \textit{monolithic} mechanism, where \ac{nrt} nodes exceeding their quota are forced into the throttle state till the completion of the \ac{rt} node: it serves as a baseline reference for \textit{ROSGuard} interval-based regulation. 
It is important to note the difference between the two methods: the thresholds considered for monolithic approach are set as a percentage over the bandwidth available over the full execution of the \ac{rt} application and, hence, are in absolute terms higher than those considered in 
the standard, interval-based 
configuration, as illustrated in Figure~\ref{fig:control_policies}. 

While the relative value is the same, the exact value for the threshold is therefore scaled according to the ratio between control interval and and full \ac{rt} node execution window.
In all the results, when determining relative bandwidth thresholds, reference values have been empirically derived from observing the \textit{total measurable bandwidth} with the most memory demanding benchmarks.

We report here the results obtained by varying the relative bandwidth threshold on two illustrative examples selected from the set of synthetic and real-world benchmarks: \textit{bandwidth\_read} and \textit{mser}, respectively representative of overdemanding and moderate applications in terms of bandwidth requirements. In the experiments, the same benchmark is deployed both as \ac{rt} and \ac{nrt} applications. Also, the self-sampling setup is considered as it generates the higher overhead on \ac{nrt} nodes, as shown in Figure~\ref{fig:pmc_impact}.

\begin{figure}
     \centering
    \includegraphics[width=1\linewidth, keepaspectratio]{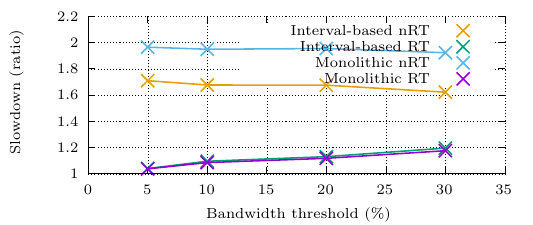}\vspace{-0.3cm}
    \caption{Slowdown ratios for two instances of \textit{bandwidth\_read}, measured when scheduled to run in parallel, at various bandwidth-thresholds. Sampling period and regulation period set to $1$ and $5$ $ms$, respectively. Monolithic regulation is reported as a reference.}
    \label{fig:threshold_isolbench}
    \vspace{-1em}
\end{figure}

\begin{figure}
     \centering
    \includegraphics[width=1\linewidth, keepaspectratio]{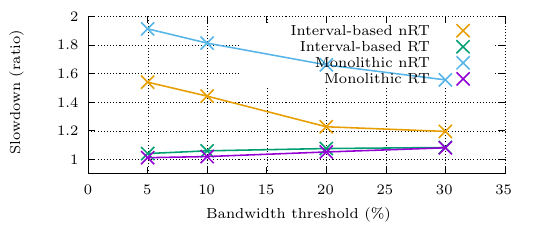}\vspace{-0.3cm}
    \caption{Slowdown ratios for two instances of \textit{mser}, 
    executed in parallel, at various bandwidth-thresholds. Sampling period and regulation period set to $1$ and $5$ $ms$, respectively. Monolithic regulation is reported as a reference.}
    \label{fig:threshold_mser}
    \vspace{-1em}
\end{figure}

Results for the two setups are reported in Figure~\ref{fig:threshold_isolbench} and~\ref{fig:threshold_mser}. We gather results for various bandwidth-threshold ratios from 5\% to 30\%, using interval-based and monolithic control policies. Self-sampling rate is set to $1ms$, while regulation period (and bandwidth budget replenishment) is fixed to $5ms$, which was determined in relation to the execution time of the considered benchmarks.

As expected, in both setups, the lower the threshold the faster it is consumed by the \ac{nrt} node, resulting in better protection and higher guarantees for the \ac{rt} application. On the downside, the \ac{nrt} nodes are forced earlier into a throttle state, negatively affecting their execution-time.
Conversely, a looser threshold is less penalizing for the \ac{nrt} applications but would lead to weaker protection from interference for \ac{rt} applications.
Results also confirm that interval-based regulation is  preferred over the monolithic approach: the latter is over penalizing the \ac{nrt} node while the former only negligibly impacts \ac{rt} nodes, still providing better overall throughput. 

In all experiments, a saturation-effect is observed on execution-times, at the lower and higher ends of the threshold-setpoint. Below certain value (10\%), the setpoint is so low that the regulation is fired \textit{instantaneously} (worst-case after just $1$ sampling period). This leads \ac{rt} applications to run near their isolation execution-time, while \ac{nrt} applications suffer considerable slowdowns (left-most values in Figures~\ref{fig:threshold_isolbench} and \ref{fig:threshold_mser}). Analogously, beyond a given value of the bandwidth-threshold, protection on the \ac{rt} applications will decrease to the point of never triggering the regulation mechanism (not shown in the plots for cleanliness).

\begin{figure}[tbp]
     \centering
    \includegraphics[width=1\linewidth, keepaspectratio]{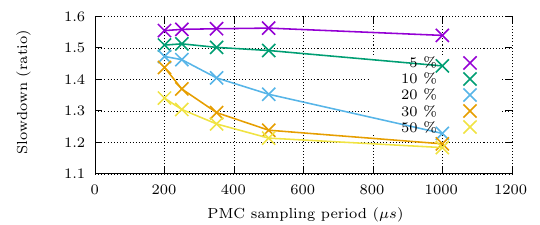}\vspace{-0.3cm}
    \caption{Slowdown ratios suffered by \textit{mser} when deployed as \ac{nrt} application, in parallel with another instance of \textit{mser} as \ac{rt} node at various bandwidth-thresholds, varying the sampling periods.}
    \vspace{-1em}
    \label{fig:impact_on_nct}
\end{figure}

Comparing the two reference benchamrks, we observe that \textit{bandwidth\_read} saturates the available bandwidth even with high (30\%) bandwidth allowance. Instead, \textit{mser} provides a more realistic example of bandwidth requirements.
For this reason, an additional experiment was conducted on \textit{mser} to analyse the impact of the sampling and regulation period on slowdown under the same threshold values. For this experiment, regulation interval is set to $5x$ the sampling period. Results, reported in Figure \ref{fig:impact_on_nct},
confirm that slowdown is also affected by the sampling-regulation period, with looser sampling-regulation leading to lower overheads on \ac{nrt} nodes, but only for sufficiently high bandwidth allowances (>10\%). 
The impact of the regulation period on \ac{rt} and \ac{nrt} nodes is further addressed in the next set of experiments.

\begin{figure}[t]
     \centering
    \includegraphics[width=1\linewidth, keepaspectratio]{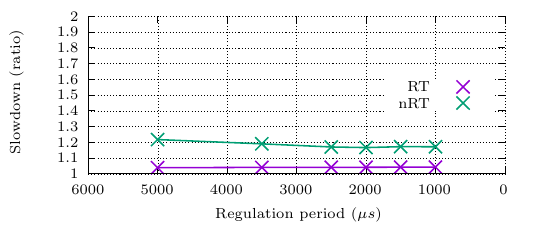}\vspace{-0.3cm}
    \caption{Slowdown ratios for two instances of \textit{mser}, measured when scheduled to run in parallel, with interval-based regulation at $30~\%$. The sampling period is set to $500$ $\mu s$.}
    \label{fig:budgetrep_mser}
    \vspace{-1em}
\end{figure}

\begin{figure}[t]
     \centering
    \includegraphics[width=1\linewidth, keepaspectratio]{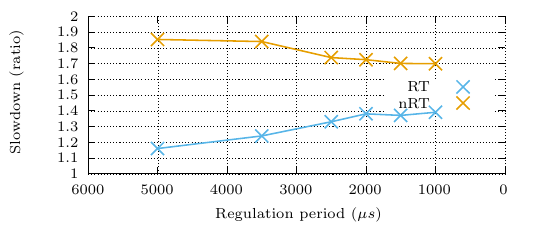}\vspace{-0.3cm}
    \caption{Slowdown ratios for two instances of \textit{bandwidth\_read}, measured when scheduled to run in parallel, with interval-based regulation at $30~\%$. The sampling period is set to $500$ $\mu s$.}
    \label{fig:budgetrep_isolbench}
    \vspace{-1em}
\end{figure}

\subsection{Impact of the Regulation Period}
This set of experiments aims to explore the effect that the regulation period has on effectiveness of \textit{ROSGuard} and the incurred overheads. For this set of experiments, we focused on \textit{ROSGuard} self-sampling configuration, as it is the bottom line in terms of \ac{nrt} nodes' overheads. 
We fixed the sampling period to $500~\mu s$ and bandwidth threshold to 30\% to isolate the impact of the regulation period. Thresholds are proportionally scaled to match the size of the regulation interval. We vary the regulation period, i.e., the frequency at which bandwidth budget in \ac{nrt} nodes is replenished, in between $1$ and $5~ms$.

\begin{figure*}[htb]
\begin{subfigure}
 \centering
    \includegraphics[width=1\linewidth, keepaspectratio]{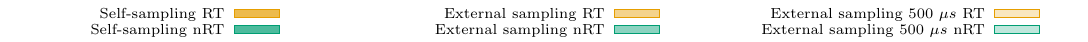}
\end{subfigure}\vspace{-0.3cm}
\begin{subfigure}
         \centering
    \includegraphics[width=1\linewidth, keepaspectratio]{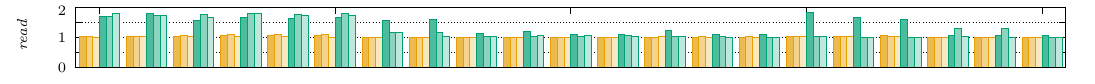}
\end{subfigure}\vspace{-0.25cm}
\begin{subfigure}
    \centering
    \includegraphics[width=1\linewidth, keepaspectratio]{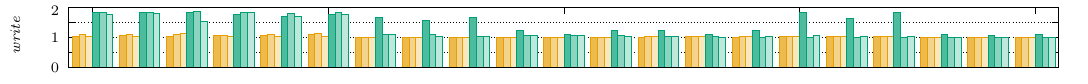}
\end{subfigure}\vspace{-0.25cm}
\begin{subfigure}
     \centering
    \includegraphics[width=1\linewidth, keepaspectratio]{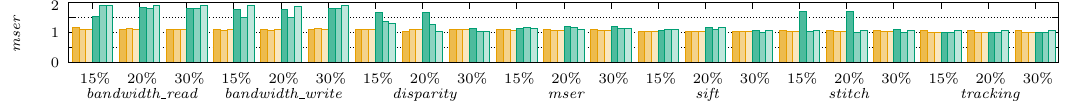}
\end{subfigure}
    \caption{Slowdown ratios measured when running various benchmark combinations with interval-based regulation at $15\%$, $20\%$ and $30\%$. Benchmarks on the vertical axis are run as \ac{rt} applications and benchmarks on the horizontal axis as \ac{nrt} applications. In the vertical axis, \textit{read} and \textit{write} refer to \textit{bandwidth\_read} and \textit{bandwidth\_write}, respectively. Colored bars indicate slowdown ratio with respect to unregulated execution-time in isolation. For each scenario we report the average slowdown for the \ac{rt} and \ac{nrt} applications, under three different configurations of \textit{RosGuard}: self-sampling and external sampling with sampling and regulation periods set to $1$ and $5$ $ms$, respectively, and an additional configuration with external sampling with sampling rate reduced to $500~\mu s$.}
    \label{fig:general_ev}
    \vspace{-1em}
\end{figure*}

\begin{figure}[htbp]
\vspace{-0.4em}
\begin{subfigure}
         \centering
    \includegraphics[width=1\linewidth, keepaspectratio]{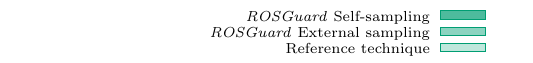}
\end{subfigure}\vspace{-.4cm}
\begin{subfigure}
     \centering
    \includegraphics[width=1\linewidth, keepaspectratio]{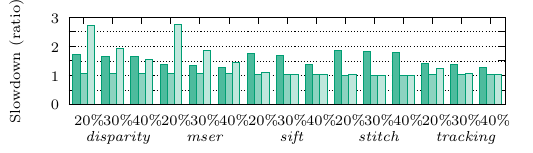}
\end{subfigure}
     \caption{Comparison of slowdown ratios to those reported for the \textit{MemGuard} implementation in \cite{izhbirdeev2024memcore} ($1000~\mu s$). Reported values include benchmarks from the \ac{sdvbs} suite, scheduled with \textit{bandwidth\_read} as \ac{rt} node,
     under $20\%$, $30\%$ and $40\%$ bandwidth thresholds, for \textit{ROSGuard} self- and external-sampling. Sampling and regulation periods are $200$ and $1000~\mu s$, respectively.
     }
     \vspace{-1.5em}
    \label{fig:memcore_comp}
\end{figure}

Figures~\ref{fig:budgetrep_mser} and \ref{fig:budgetrep_isolbench} show the slowdown ratios of \ac{rt} and \ac{nrt} applications when varying the regulation interval for \textit{mser} and \textit{bandwidth\_read} benchmarks (both as \ac{rt} and \ac{nrt} node). 
In general, by increasing the frequency of regulation, the \ac{nrt} nodes' budget is replenished more frequently, possibly allowing them to incur fewer interruptions and ultimately accumulate less delay. Conversely, the protection enforced on the {green}\ac{rt} node becomes less robust. The actual impact, however, is application-dependent as it depends on the distribution of bandwidth requirements and on how close those requirements are to the reserved bandwidth quota. This is indeed confirmed by our experiments as for 
\textit{mser}, the benchmark with lower bandwidth requirements, we observe minimal variation, whereas 
for \textit{bandwidth\_read}, the benchmark with higher bandwidth requirements (>30\% as observed in Figure~\ref{fig:threshold_isolbench}), we observe that \ac{nrt} node is penalized by higher regulation intervals and vice-versa.

\subsection{Impact of the Application's Bandwidth Profile}\label{sec:general_ev}
As observed above, the impact of the regulation mechanism on application timing is application dependent and largely varies depending on the \ac{rt} and \ac{nrt} applications' bandwidth requirements. 
With the next extensive set of experiments, we aim at characterizing the sensitivity of the overhead to applications' bandwidth profile by assessing the overall performance of \textit{ROSGuard} when facing different combinations of benchmarks as \ac{rt} and \ac{nrt} nodes. 
As \ac{rt} applications we considered \textit{mser}, \textit{bandwidth\_read}, and \textit{bandwidth\_read}, while as \ac{nrt} nodes we deployed all the considered benchmarks.
The characterization of the benchmarks is provided at the beginning of this Section.  

We consider both sampling configurations in \textit{ROSGuard} (self- and external sampling variants): while protection from timing-interference is adequate for all applications, external sampling is less penalizing for \ac{nrt} applications and guarantees more rapid reaction in case of threshold exceedance.

In Figure~\ref{fig:general_ev}, we report the average relative slowdown incurred by the \ac{rt} and \ac{nrt} nodes with respect to unregulated execution in isolation.  
We configured \textit{ROSGuard} with a sampling period of $1~ms$ and a regulation period of $5~ms$, which provides a good trade-off between accuracy and overhead. Additionally, for the external sampling setup only, we deploy a configuration with a higher sampling frequency, at $500~\mu s$.
In order to include the impact of bandwidth threshold across all benchmark combinations, we considered different threshold values at $15\%$, $20\%$ and $30\%$. \textit{ROSGuard} achieves good results in all considered combinations, guaranteeing performance for the \ac{rt} nodes that are pretty close to performance in isolation, both under the self and external sampling schemes. 
Excluding high-bandwidth benchmarks (\textit{bandwidth\_read}, \textit{bandwidth\_write}, and \textit{disparity}), the impact on \ac{nrt} nodes is limited.

On the accuracy side, with a sampling period of $1~ms$, the difference between self- and external-sampling schemes is marginal. By increasing the sampling rate to $500~\mu s$, the protection mechanism generally produces a positive impact on \ac{rt} performance, hinting at an increased precision in the regulation. 
In some cases, however, the increased granularity offered by the external sampling has the counter intuitive effect of
causing further delays in the \ac{rt} node. This scenario, already discussed in~\cite{brilliWCET24}, can be observed for the configurations involving, the high-bandwidth benchmarks from Isolbench (\textit{bandwidth\_read}, and \textit{bandwidth\_write}) as \ac{nrt} nodes, which are causing an exceptionally high bandwidth saturation.

On \ac{nrt} nodes, the external sampling scheme generally exhibits better performance due to the reduced overheads.
For the benchmarks from Isolbench (\textit{bandwidth\_read}, and \textit{bandwidth\_read}), their bandwidth requirements are already so much beyond the $30\%$ of the available bandwidth that sampling and regulation configuration cannot produce a substantial difference. With a constantly high bandwidth saturation (at $30\%$), the fact that the \ac{nrt} application is inhibited slightly earlier thanks to the lower activation delay guaranteed by the external sampling configuration, may sometimes accumulate into a non-negligible increase in execution time. 

The fact that \textit{ROSGuard} addresses different regulation granularities than those of reference approaches, makes direct comparison complicated. To enable a reasonable comparison, we compare our results under a configuration that is compatible with that of \textit{Memguard} reported in~\cite{zuepke2023mempol} (with a $1~ms$ regulation) on a subset of \textit{SD-VBS}.
We set the regulation period to $1~ms$ (same as \textit{MemGuard} in~\cite{zuepke2023mempol}) and consider the same bandwidth thresholds of $20\%$, $30\%$ and $40\%$. \textit{ROSGuard} sampling rate is configured at $200~\mu s$. The difference in the setup and platform allows only a coarse-grained assessment but, in general, results reported in Figure~\ref{fig:memcore_comp} are comparable.
First of all, by comparing \textit{ROSGuard} results with those in Figure~\ref{fig:general_ev}, we observe the higher overhead incurred by \ac{nrt} nodes due to the higher sampling rate.
For \textit{disparity} and \textit{mser}, the apparent improvement of \textit{ROSGuard} over \textit{MemGuard} is probably justified by the overhead being compensated by a higher available bandwidth on the AGX Orin. Results on other benchmarks, instead, confirm that \textit{ROSGuard} sampling overhead makes it more amenable to looser bandwidth regulation requirements. The external sampling configuration seems to remove the sampling overhead  
but we are aware that the comparison with \textit{MemGuard} is not totally fair as results heavily depend on the total available bandwidth in the target.
\section{Conclusion}\label{sec:conclusion}
The solutions proposed in the literature for bandwidth regulation mechanisms are generally addressing strongly constrained setups where regulation must be applied at extremely fine granularity, typically requiring bandwidth monitoring to happen at frequencies in the $\mu s$ order. This is ultimately over the demands of a wide class of ROS2-based applications with looser real-time constraints, but still requiring protection from unregulated interference. 
We designed and implemented \textit{ROSGuard}, a fully functional bandwidth regulation mechanism, 
that provides a highly configurable and portable solution that better fits the control granularity required by \ac{ros2}-based applications. \textit{ROSGuard} effectively protects critical nodes from timing interference, while at the same time guaranteeing low overheads and high memory throughput. As future work, we plan to extend the set of configurable regulation policies and further improve the control on \ac{ros2} execution over the OS layer.


\bibliographystyle{IEEEtran}
\bibliography{bibs/1_introduction,
                bibs/2_background,
                bibs/4_evaluation
                }
\end{document}